\documentclass[superscriptaddress,aps,prl,twocolumn,showpacs,nofootinbib, longbibliography, notitlepage,floatfix]{revtex4-2}
\usepackage{etex}
\usepackage{amsmath,amssymb,amsthm}
\usepackage[colorlinks=true,citecolor=blue,urlcolor=blue]{hyperref}
\usepackage[pdftex]{graphicx}
\usepackage{times,txfonts}
\usepackage{braket}
\usepackage{color}
\usepackage{natbib}
\usepackage{amsmath,blkarray}
\usepackage{mathtools}
\usepackage{ulem}
\usepackage{latexsym}
\usepackage{tabularx, booktabs}
\usepackage{graphics,epstopdf}
\usepackage{graphicx}
\usepackage{float}
\usepackage{amsfonts}
\usepackage{tikz}
\usetikzlibrary{quantikz}
\usepackage{color,soul}
\usepackage{lineno}

\newcommand{\be}{\begin{equation}}
\newcommand{\ee}{\end{equation}}
\newcommand{\ba}{\begin{eqnarray}}
\newcommand{\ea}{\end{eqnarray}}

\newcommand{\tr}{\operatorname{Tr}}

\usepackage{multirow}
\usepackage{appendix}
\usepackage{url}

\newcommand{\expv}[2]{\left\langle{#1}\right\rangle_{#2}}

\begin{document}


\title{ NMR Protocol for Black-Box Ergotropy Estimation via Feedback Algorithm }

\author{Jitendra Joshi}
\email{jitendra.joshi@students.iiserpune.ac.in}
\affiliation{Department of Physics and NMR Research Center, Indian Institute of Science Education and Research, Pune 411008, India}

\author{T. S. Mahesh}
\email{mahesh.ts@iiserpune.ac.in}
\affiliation{Department of Physics and NMR Research Center, Indian Institute of Science Education and Research, Pune 411008, India}

\begin{abstract}
Considering the emerging applications of quantum technologies, studying energy storage and usage at the quantum level is of great interest.  In this context, there is a significant contemporary interest in studying \textit{ergotropy}, the maximum amount of work that can be extracted unitarily from an energy-storing quantum device.
Here we propose a feedback-based quantum algorithm for ergotropy estimation using Lyapunov control techniques. 
By iteratively adjusting the strengths of applied drive fields, our algorithm achieves both unitary energy extraction and passive state preparation efficiently. Unlike previous approaches, this algorithm does not require any classical optimization, is resilient to cumulative errors, and eliminates the need for any time-consuming quantum state tomography. We  validate these advantages via numerical simulations demonstrating robustness even under noisy conditions. Additionally, we experimentally implement the algorithm on up to three-qubit NMR registers, establishing its practical viability.

\end{abstract}

\maketitle
Quantum thermal machines, such as refrigerators and heat engines, play a pivotal role in exploring the principles of quantum thermodynamics by regulating heat flow and work production~\cite{kosloff2014quantum,mitchison2019quantum}. Complementary to these, quantum batteries offer a promising avenue for energy storage and extraction~\cite{alicki2013entanglement,hovhannisyan2013entanglement,rossini2019many,andolina2019extractable,rodriguez2024optimal,hu2022optimal,konar2022quantum,barra2022quantum,campaioli2024colloquium,campaioli2018quantum,binder2015quantacell,shastri2024dephasing}. Exhibiting advantages over classical systems, quantum batteries demonstrate superior performance in key metrics, including faster charging speeds~\cite{andolina2019quantum,ghosh2020enhancement,gao2022scaling,gyhm2022quantum,salvia2023quantum,rodriguez2023catalysis,PhysRevA.106.042601}, higher energy storage~\cite{zhang2019powerful,andolina2018charger,PhysRevA.106.042601}, and enhanced energy extraction capabilities~\cite{ferraro2018high,le2018spin,tirone2023work,liu2021entanglement,monsel2020energetic,chaki2024positivenonpositivemeasurementsenergy,chaki2023auxiliaryassistedstochasticenergyextraction}.

Extracting maximum energy from quantum systems via unitary processes, known as ergotropy~\cite{allahverdyan2004maximal}, is a fundamental task in thermodynamics. For a quantum system in state $\rho_0$ and the system Hamiltonian $H_0$, ergotropy can be formally defined as 
\begin{align}
{\cal E}(\rho_0)=E(\rho_0)-E(\rho^p),
\label{eq:ergo}
\end{align}
where $E(\rho^p) = \min_{U} \tr\left(U\rho_0 U^{\dagger}H_0\right)$. The unitary ${\cal U}_p$ minimizes the system's energy, and the passive state is expressed as $\rho^p = {\cal U}_p\rho_0 {\cal U}_p^{\dagger}$~\cite{allahverdyan2004maximal,pusz1978passive}. The task of extracting energy is complex due to the sensitivity of ergotropy to quantum correlations, which can significantly impact performance~\cite{goold2016role,bera2017generalized,manzano2018optimal,vitagliano2018trade,gemme2022ibm,dou2022cavity,dou2022extended,dou2023superconducting}. Recent studies have explored the interplay between entanglement and thermodynamic quantities, such as the ergotropy gap, which quantifies the difference in maximum work extractable via global and local unitaries. These quantities have been used as tools to verify entanglement in bipartite~\cite{PhysRevA.99.052320}, multipartite~\cite{PhysRevA.109.062427,PhysRevLett.129.070601}, and multiqubit mixed states~\cite{PhysRevA.109.L020403}. Experimental verification of entanglement using these measures has been demonstrated in systems with up to ten qubits~\cite{PhysRevA.109.L020403}.

A natural approach to determining the extractable energy from a quantum system is to drive the system into its lowest achievable energy state, capturing the surplus energy during this process. One key application of modern quantum technologies is finding low-energy configurations of quantum systems. This includes methods such as variational quantum algorithms~\cite{cerezo2021variational}, adiabatic quantum computing~\cite{RevModPhys.90.015002}, and others. Among these, implementing quantum Lyapunov control (QLC)~\cite{kuang2008lyapunov,hou2012optimal} techniques have recently garnered significant attention. These algorithms are particularly appealing as they avoid resource-intensive classical optimization and, under specific conditions, ensure monotonic convergence to the lowest energy state. However, while QLC-based approaches have been predominantly employed for determining ground states~\cite{PhysRevLett.129.250502,chandarana2024lyapunovcontrolledcounterdiabaticquantum,tang2024nonvariationaladaptalgorithmquantum} and excited states~\cite{rahman2024feedbackbasedquantumalgorithmexcited}, their potential applications in quantum thermodynamics remain largely unexplored.

In this letter, we introduce a feedback-based quantum algorithm for ergotropy estimation, termed FQErgo. The algorithm leverages QLC techniques to prepare passive states, enabling accurate ergotropy estimation. FQErgo presents several advantages over existing methods. Firstly, its fully quantum nature eliminates the need for classical optimizers, distinguishing it from previous approaches~\cite{PhysRevResearch.6.013038,PhysRevLett.130.210401}, and allows for seamless automation in practical implementations. Secondly, FQErgo is inherently robust against circuit errors, as each feedback iteration dynamically adjusts the drives to mitigate prior inaccuracies, preventing cumulative error propagation. Lastly, circuit parameters are efficiently measured using an ancillary probe qubit via the interferometric method~\cite{moussa2010testing,PhysRevA.90.022303,mahesh2015ancilla}, avoiding the computational overhead of quantum state tomography. We validate the algorithm through numerical simulations and experimentally demonstrate its implementation on two- and three-qubit NMR systems.

\begin{figure}
\centering
\includegraphics[trim={0cm 0cm 0cm 0cm},clip,width=6cm]{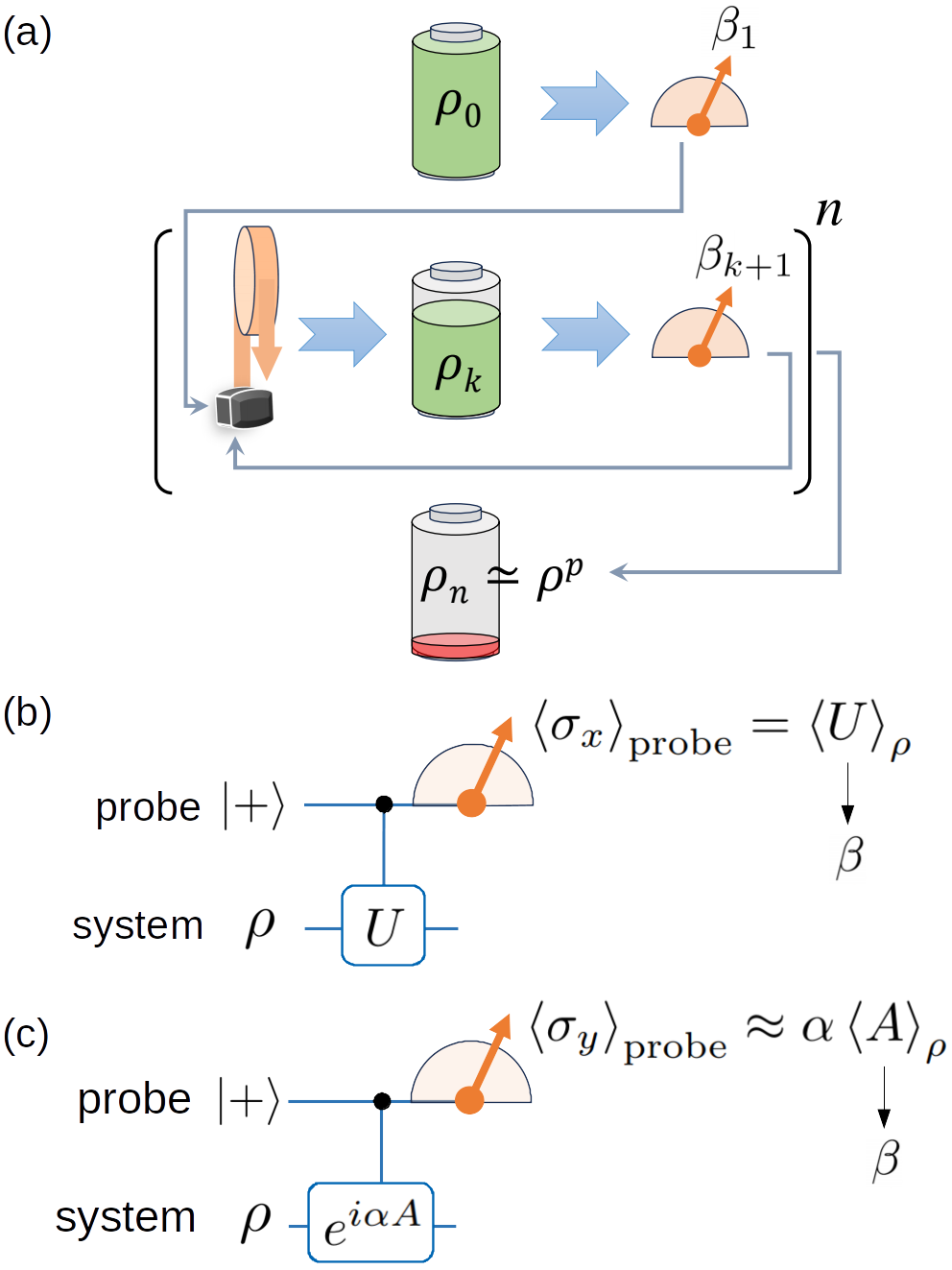}
\caption{\label{fig:FQErgo}
(a) Illustrating $n$ iterations of FQErgo applied on an initial state $\rho_0$ to reach $\rho_n \approx \rho^p$, its passive state. (b,c) Determining FQErgo drive amplitude $\beta$ with a probe qubit, for  (b) unitary Hermitian drive $\sigma_\gamma$, and  (c) nonunitary Hermitian drive $A$.  For energy measurement, we set $A=H_0$.}
\end{figure}

{\it FQErgo:} 
Consider a scenario where multiple copies of individually accessible quantum systems are prepared in an identical but potentially unknown state $\rho_0$. Alternatively, an oracle may be assumed to transform a uniquely initialized quantum system into a well-defined but unknown state $\rho_0$. To manipulate the system, a drive Hamiltonian of the form $\beta(t) H_d$ is introduced, where $\beta(t)$ represents a time-dependent control parameter. The system dynamics are governed by 
\begin{equation}
i\frac{d}{dt}\rho(t) = \left[H_0+\beta(t)H_d , \rho(t)\right],
\end{equation}
with $\hbar = 1$. To minimize the system's energy, $E(\rho(t)) =\expv{H_0}{\rho(t)} = \tr(H_0 \rho(t))$, the control parameter $\beta(t)$ is designed to satisfy the Lyapunov condition 
\begin{equation}
\frac{d}{dt}\expv{H_0}{\rho(t)} = \beta(t) \, \expv{C_d}{\rho(t)} \le 0, ~~ \forall~ t \ge 0
\end{equation}
where $C_d = i[H_d, H_0]$. This is achieved by setting $\beta(t) = -w \, \tr(C_d \rho(t))$, with $w$ as a positive scalar coefficient \cite{PhysRevLett.129.250502}. 

Hence, the algorithm proceeds as follows. In the first step, the control parameter $\beta_1 = -w \, \tr(C_d \rho_0)$ is calculated by measuring the expectation value $\tr(C_d \rho_0)$. For the first iteration, the state is then updated according to $\rho_1 = U_1 \rho_0 U_1^\dagger$ where $U_n = u_n  \cdots u_2  u_1$ , with $u_k = e^{-i\beta_k \tau H_d }$. For subsequent iterations, the control parameter is determined iteratively as $\beta_k = -w \, \tr(C_d \rho_{k-1})$, with the updated state prepared as $\rho_k = U_k \rho_{k-1} U_k^\dagger$. By repeating this iterative process for $n$ steps, the system evolves from the initial state $\rho_0$ to a final state $\rho_n \approx \rho_p$, where $\rho_p$ is the passive state that minimizes the system's energy. The energy difference between the initial state $\rho_0$ and the final state $\rho_n$ represents the ergotropy, as quantified by Eq.~\eqref{eq:ergo}.

The implementation of FQErgo relies on efficiently evaluating $\expv{C_d}{\rho(t)}$ and monitoring the system energy without resorting to quantum state tomography, which is computationally expensive and unnecessary. To achieve this, we utilize an interferometric circuit with an ancillary probe qubit \cite{moussa2010testing,PhysRevResearch.6.013038,PhysRevLett.130.210401}. As shown in Fig.~\ref{fig:FQErgo}(b,c), the circuit measures the energy $E(\rho_k)$ and extracts the drive parameter $\beta_k$. For a Hermitian observable $U$, the expectation value $\expv{U}{\rho}$ can be obtained by initializing the probe qubit in the $\ket{+}$ state and applying a controlled-$U$ gate. The resulting probe signal is $\expv{\sigma_x}{\text{probe}} = \expv{U}{\rho}$ \cite{PhysRevA.90.022303}, providing the desired expectation value. For example, if the system Hamiltonian is $\sigma_z$, the probe directly measures $\expv{\sigma_z}{\rho}$. To extract $\expv{A}{\rho}$ for a nonunitary Hermitian observable $A$, we construct the unitary operator $e^{-i \alpha A}$, where $\alpha$ is a small real parameter. Using the circuit in Fig.~\ref{fig:FQErgo}(c), the probe signal becomes $\expv{\sigma_y}{\text{probe}} = -\expv{\sin(\alpha A)}{\rho} \approx -\alpha \expv{A}{\rho}$ \cite{PhysRevA.90.022303}. Setting $A = C_d$, the expectation value $\expv{C_d}{\rho}$ is directly obtained by measuring $\expv{\sigma_y}{\text{probe}}$. This approach is especially advantageous when the operator to be measured is non-diagonal in the measurement basis. It reduces the measurement overhead compared to conventional methods that require a change of basis to measure the operator but at the cost of introducing an ancillary qubit. For experiments with larger system sizes the trade-offs between different methods must be considered.

\begin{figure}
\centering
\includegraphics[trim={0cm 0cm 0cm 0cm},clip,width=9cm]{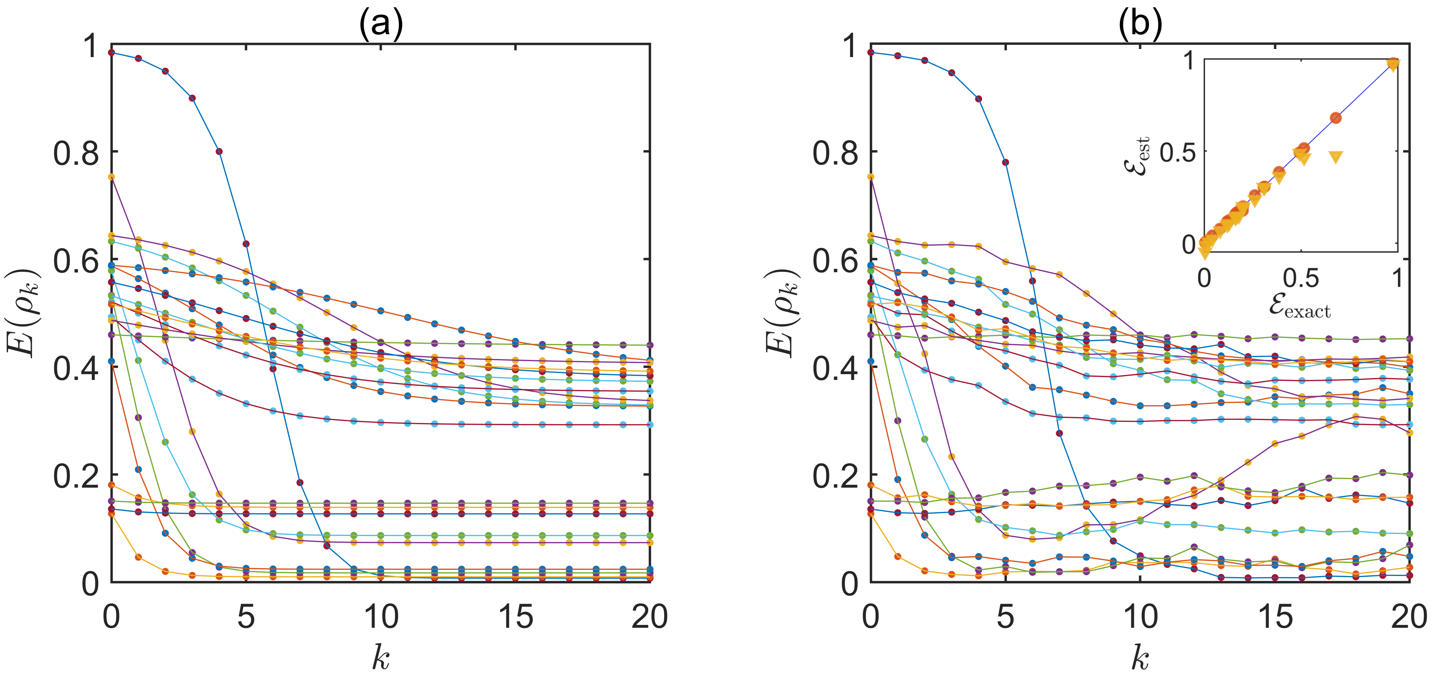} 
\caption{(a,b) Simulated system-energy $E(\rho_k)$ vs iteration number $k$ for a single qubit ($\omega_0=1)$ starting from 20 random initial states applied with  (a) ideal FQErgo and  (b) FQErgo having random errors.  Inset figure shows numerically estimated vs exact ergotropy values with ideal FQErgo (filled circles) and FQErgo with random errors (filled triangles). 
\label{fig:1qubit}}
\end{figure}


\begin{figure}
\centering
\includegraphics[trim={0cm 0cm 0cm 0cm},clip,width=8cm]{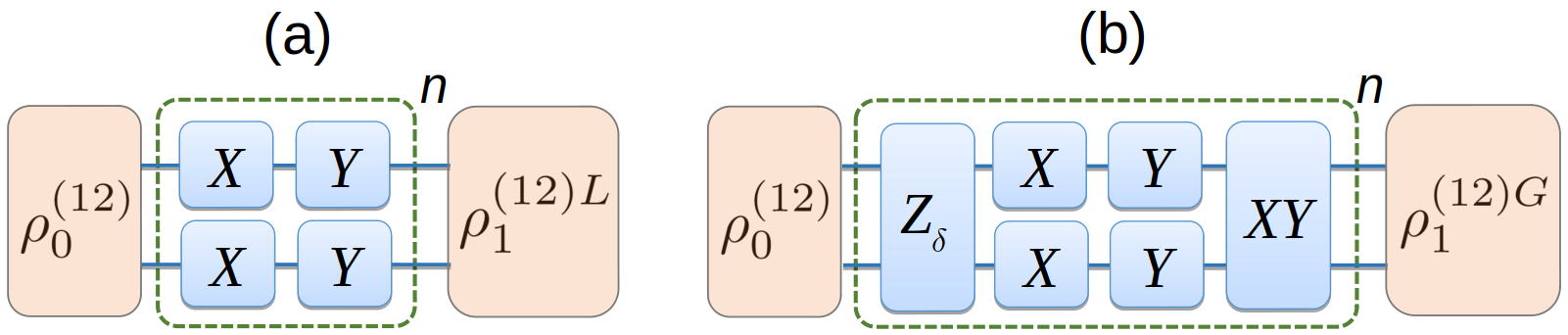} \\
\includegraphics[trim={2cm 0cm 2cm 0cm},clip,width=8.7cm]{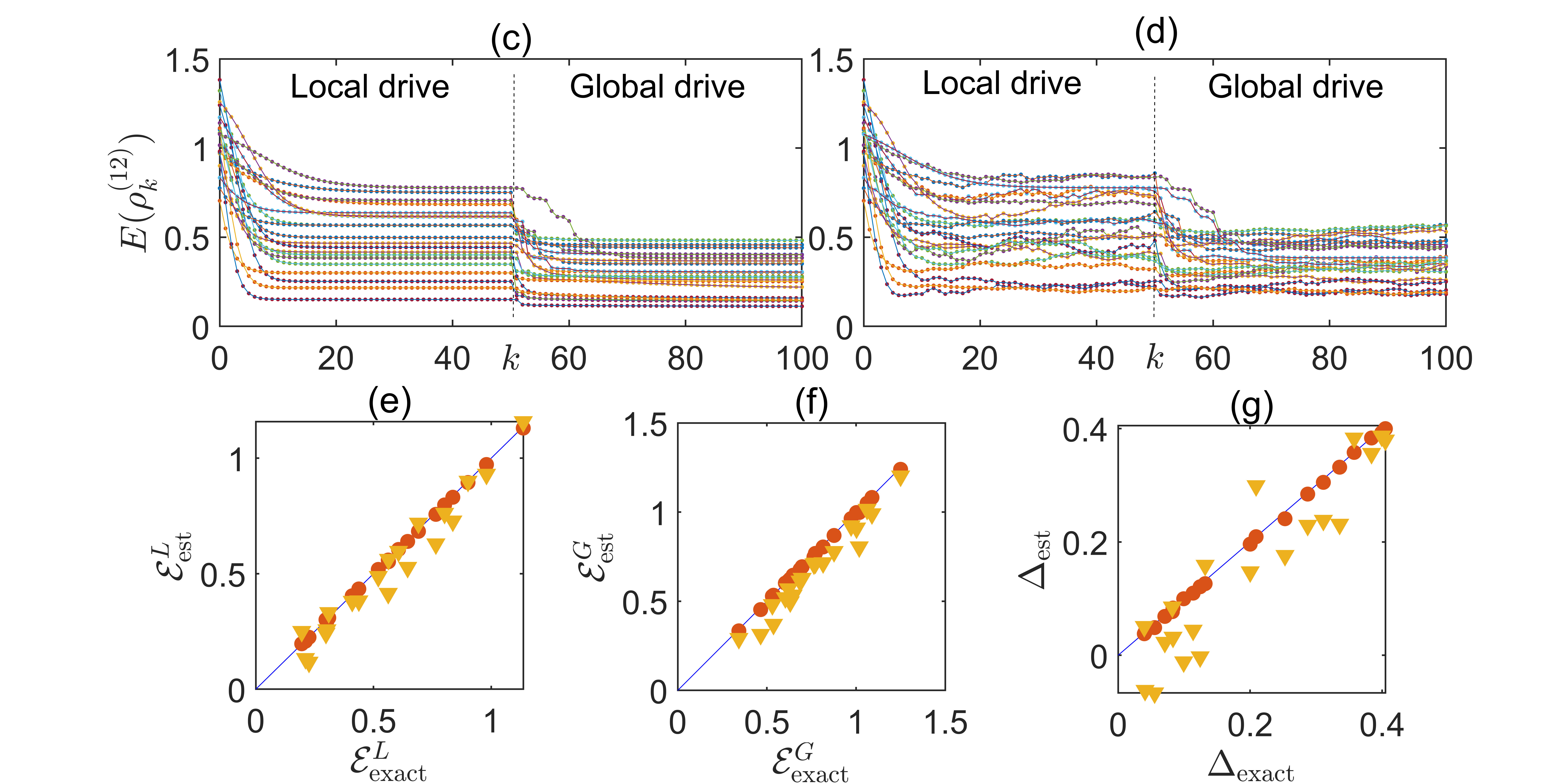} 
\caption{\label{fig:ergo_twoqubit}
(a,b) FQErgo circuits for local (a) and global (b) energy extraction in two-qubits.  Here, local drives $X$, $Y$ suffice for reaching the local passive state $\rho^{(12)L}$ and estimating the local ergotropy ${\cal E}^L(\rho^{(12)}_0)$  (a). In contrast, additional global drive such as $XY$ and $Z_\delta$ \cite{PhysRevResearch.6.013038} gates are necessary for reaching the global passive state $\rho^{(12)G}$ and estimating the global ergotropy ${\cal E}^G(\rho^{(12)}_0)$ (b).  
(c,d) Simulated system-energy vs iteration number for the two-qubit ($\omega_0 = 1$ and $J = 0.01$) FQErgo from 20 random initial states without error (c) and with error (d).  (e,f) The numerically estimated local (e) and global ergotropy (f) vs the exact values.
(g) Ergotropy gap obtained from (e) and (f), vs the exact values.  In (e-g), the filled circles are without random errors and filled triangles are with random errors.}
\end{figure}

{\it Numerical simulations:} To evaluate the performance of FQErgo, we consider a Hamiltonian $H_0 = \omega_0(\mathrm{I}-\sigma_z)/2$, with energy eigenvalues $(0, \omega_0)$, and an arbitrary mixed state $\rho(0) = (1-\epsilon)/2 \, \mathrm{I} + \epsilon \proj{\psi}$, where $\ket{\psi} = \cos(\theta/2)\ket{0} + e^{i\phi}\sin(\theta/2)\ket{1}$ and the purity parameter $\epsilon \in [0,1]$. FQErgo is implemented to evolve the state to its passive form, $\rho^p = (1-\epsilon)\mathrm{I}/2 + \epsilon \proj{0}$. In this setup, we select drive $H_{d} =  \sum_i\sigma_x^i+\sum_i\sigma_y^i$ to reach $\rho^p$ efficiently with ergotropy ${\cal E}_\mathrm{est}(\rho_0) = E(\rho_0) - E(\rho_n)$. Figure~\ref{fig:1qubit}(a) illustrates the energy $E(\rho_k)$ as a function of the iteration number $k$ for $20$ random initial states. In all cases, the energy decreases monotonically, converging to the respective passive states $\rho^p$ after sufficient iterations. In Fig.~\ref{fig:1qubit}(c), a comparison of ${\cal E}_\mathrm{est}(\rho_0)$ with ${\cal E}_\mathrm{exact}(\rho_0) = E(\rho_0) - E(\rho^p)$ confirms perfect estimation of ergotropy for all the random initial states.

We now numerically investigate the robustness of FQErgo against potential imperfections in practical implementations. A key characteristic of FQErgo is its intrinsic resilience to variations in drive amplitude, owing to its feedback-based nature. If the drive amplitude is reduced below the expected value, the algorithm still converges to the passive state, at the cost of having more iterations. To examine the impact of more general errors, we introduce a 5-degree rotation error about a random axis during each FQErgo iteration. As shown in Fig.~\ref{fig:1qubit}(b), the system-energy curves no longer display a perfectly monotonic decay but do not exhibit a significant accumulation of errors either. The corresponding ergotropy values, represented by filled triangles in Fig.~\ref{fig:1qubit}(c), show slight underestimations. This behavior is expected, as circuit errors may prevent the system from fully reaching its passive state, causing it to settle in a higher energy configuration.

It is essential to highlight a fundamental distinction between work extraction in single-qubit and two-qubit systems. For two-qubit system $\rho_{12}$, energy can be extracted through either local unitaries $U_L = U_1\otimes U_2$ acting on individual subsystems or global unitaries $U_G = U_{12}$ applied to the entire system. When the minimization of energy is restricted to local unitaries, the system reaches the local passive state $\rho^{p_L}$, which defines the local ergotropy, ${\cal E}^{L} = E(\rho_0) - E(\rho^{p_L})$. On the other hand, if global unitaries are used for minimization, the system evolves to the global passive state $\rho^{p_G}$, determining the global ergotropy, ${\cal E}^{G} = E(\rho_0) - E(\rho^{p_G}) $. The difference, $\Delta = {\cal E}^{G} - {\cal E}^{L}$, is referred to as the ergotropy gap which is always greater or equal to zero as $U_L \subseteq U_G$  ~\cite{PhysRevX.5.041011}.

We consider a two-qubit system with the Hamiltonian $H_0 = \omega_0(\mathrm{I}\otimes\mathrm{I} - (\sigma_z\otimes\mathrm{I} + \mathrm{I}\otimes\sigma_z)/2) + J(\sigma_z\otimes\sigma_z),$ where $|J| \ll |\omega_0|$, prepared in random initial states $\rho_0^{(12)}$. The ergotropy gap is defined as 
\begin{align}
\Delta = {\cal E}^G\left(\rho^{(12)}_0\right) - \left({\cal E}^L\left(\rho^{(1)}_0\right) + {\cal E}^L\left(\rho^{(2)}_0\right)\right).
\end{align}
To validate the performance of FQErgo in estimating both local and global ergotropies and, consequently, the ergotropy gap, we utilize local drives $H_{d} = \sum_{\gamma}\sum_{i=1}^2 \sigma_\gamma^i$, where $\gamma \in \{x, y\}$. The results are shown in Fig.~\ref{fig:ergo_twoqubit}(a). Over $k = 30$ iterations, the system energy saturates, reaching $\rho_{30} \approx \rho^{(12)p_L}$. From $k = 31$, a further reduction in energy is observed by combining global and local operations, as illustrated in Fig.~\ref{fig:ergo_twoqubit}(b). This includes applying a tilted phase gate $Z_\delta = e^{-i\delta (\mathrm{I}\otimes \sigma_y)} \left[ \proj{0} \otimes \mathrm{I} + \proj{1} \otimes e^{-i\sigma_z \pi/2} \right] e^{i\delta (\mathrm{I}\otimes \sigma_y)},$ with a fixed small angle $\delta$~\cite{PhysRevResearch.6.013038}. The driving Hamiltonian is set as $H_d = \sum_{i=1}^2 \sigma_x^i + \sum_{i=1}^2 \sigma_y^i + \sum_{i,j=1}^2 \left(\sigma_x^i \sigma_y^j + \sigma_y^i \sigma_x^j\right),$ with the corresponding unitary $u_k = e^{-i \beta^k H_d \tau}$, where $\beta^k$ is the drive strength and $\omega_0\tau = 0.8$. After $k = 60$ iterations, the system energy saturates again, reaching $\rho_{60} \approx \rho^{(12)p_G}$.

Figure~\ref{fig:ergo_twoqubit}(c) depicts the monotonic decrease in energy for 20 random initial states using FQErgo. For each initial state, two minimum-energy states are identified: one corresponding to the local passive state and the other to the global passive state. Figures~\ref{fig:ergo_twoqubit}(e-f) present a comparison between the estimated and exact values of local and global ergotropies, while Fig.~\ref{fig:ergo_twoqubit}(g) shows the estimated ergotropy gaps plotted against their exact values. The global energy extraction sequence in Fig.~\ref{fig:ergo_twoqubit}(b) could be employed directly from the initial state to achieve faster convergence to the passive state, but the primary focus here is on estimating the ergotropy gap $\Delta$.

We further examine the robustness of FQErgo by introducing a random nonlocal error unitary, $U_\text{err} = e^{-i H_\text{err} \eta}$, where $H_\text{err}$ is a random unit-norm error Hamiltonian and $\eta \equiv 2$ degrees, applied during each FQErgo iteration. As shown in Fig.~\ref{fig:ergo_twoqubit}(d), the system-energy curves lose monotonicity but maintain the overall decreasing trend. The error-affected local and global ergotropy values, along with the corresponding ergotropy gaps, are represented by filled triangles in Figs.~\ref{fig:ergo_twoqubit}(e-g). Similar to the single-qubit case, the ergotropy values—especially the global ergotropy—are slightly underestimated. However, the root-mean-square deviation of the estimated ergotropy gaps from the ideal values remains below 0.07, demonstrating the reasonable robustness of FQErgo against unitary errors.

\begin{figure}
\centering
\includegraphics[trim={0.3cm 0cm 0.2cm 0cm},clip,width=2.0cm]{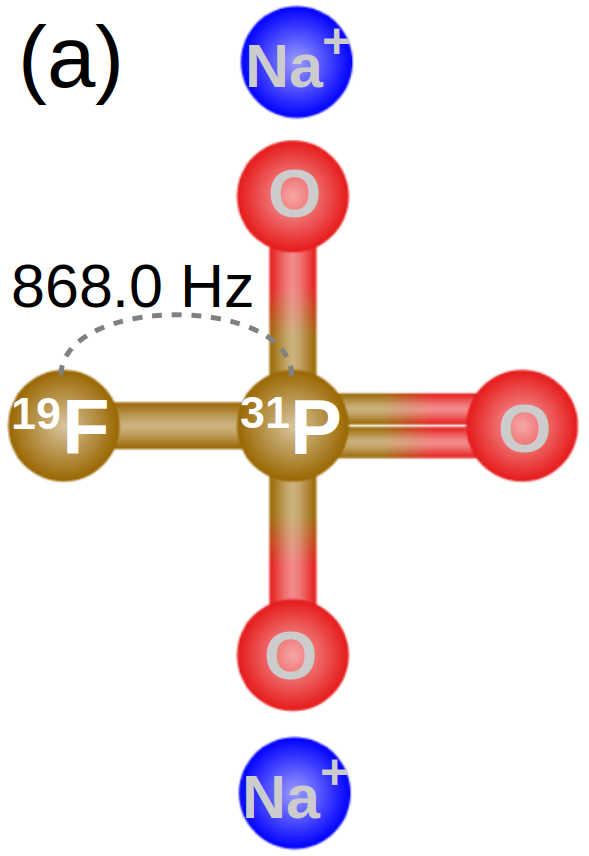}
\includegraphics[trim={0.6cm 2.5cm 5cm 4.5cm},clip,width=4.8cm]{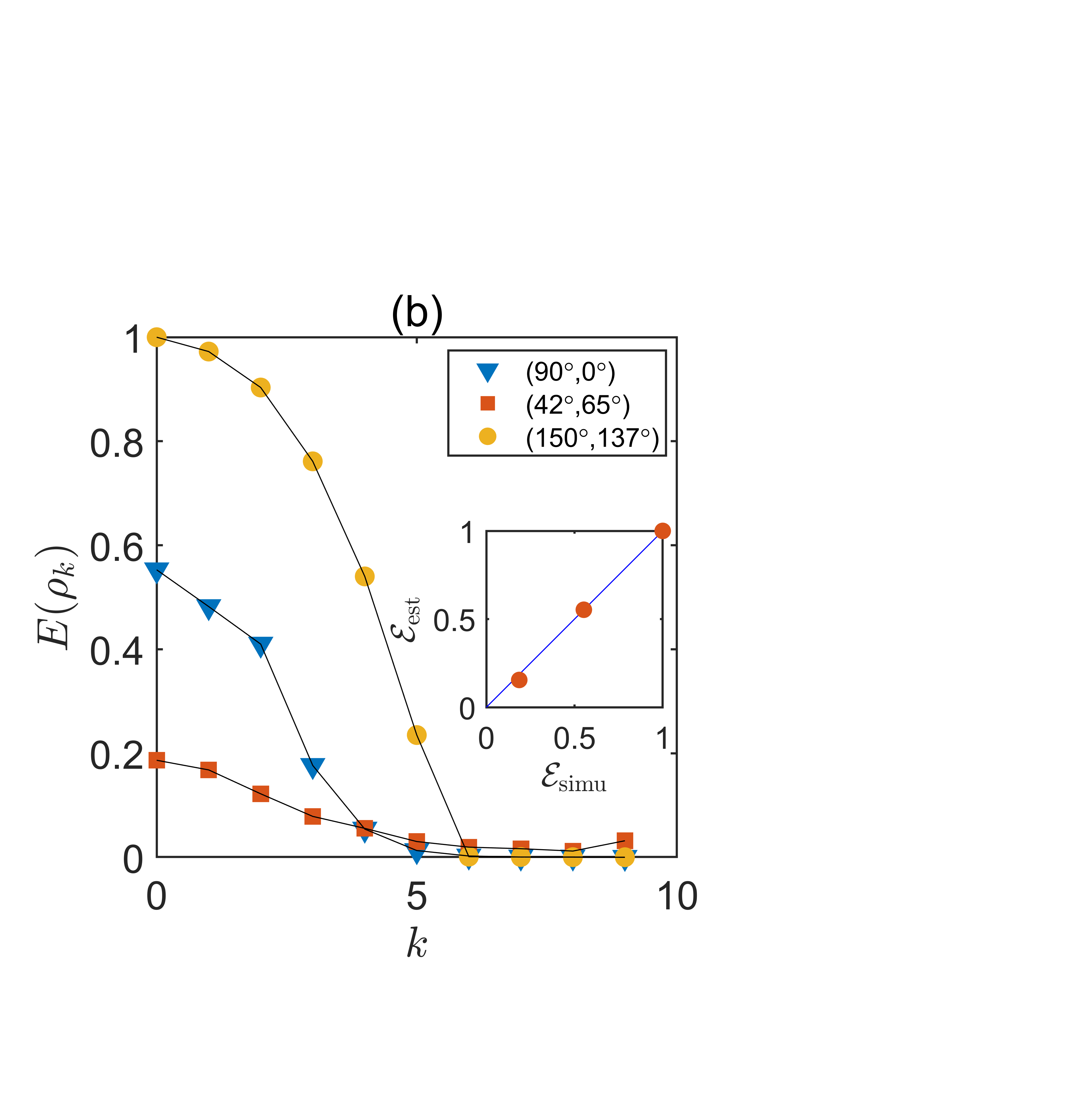}
\\
\includegraphics[trim={0cm 0cm 0cm 0cm},clip,width=9cm]{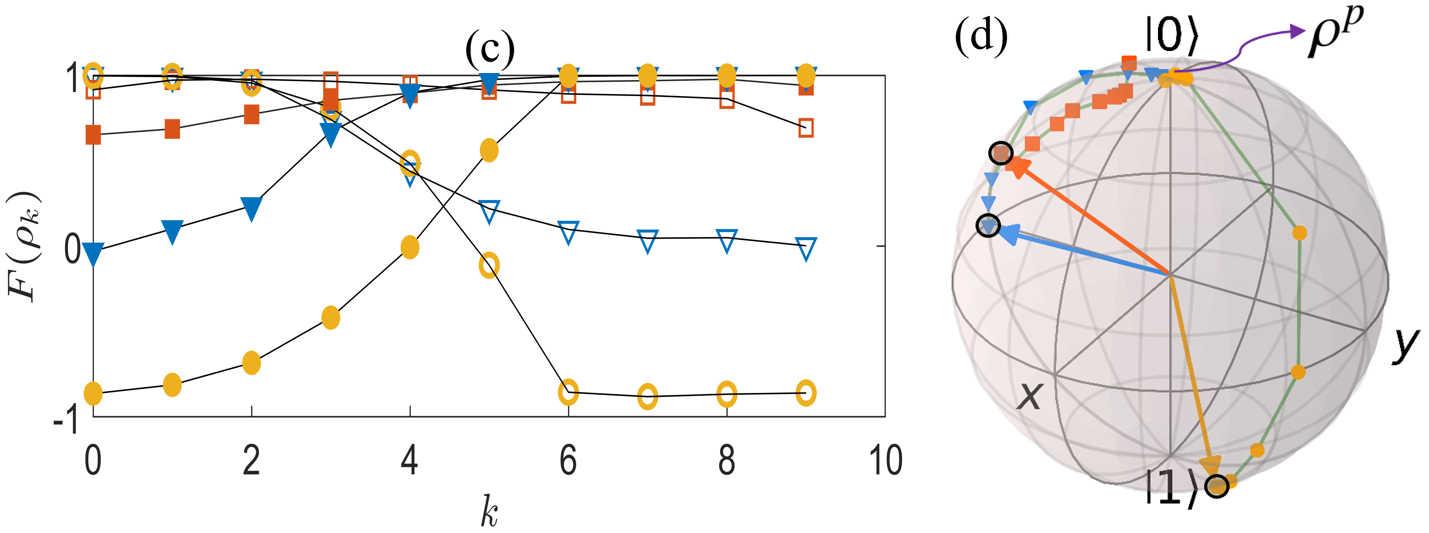}
\caption{\label{fig:1qexp}
(a) Sodium fluorophosphate molecular structure with $J_\mathrm{FP}$ indicated. (b) The experimentally measured one-qubit energy (normalized) vs FQErgo iterations for three initial states $\rho_{\theta,\phi}  = \proj{0} \otimes U_{\theta,\phi} \proj{0} U^\dagger_{\theta,\phi}$ prepared by one of the RF rotations $U_{\theta,\phi} = e^{-i\theta(I_x\cos\phi + I_y\sin\phi)} \equiv (\theta,\phi)$ as indicated.  The inset figure shows ergotropy estimated from (b) vs simulated values. (c) Decaying fidelities of initial states ($F_0(\rho_k)$, open symbols) and corresponding growth of passive states ($F_p(\rho_k)$, filled symbols) for the same three cases as in (b). (d) Bloch sphere evolution from initial state to passive state for the same three cases.}
\end{figure}

{\it Experiments:}
We utilized sodium fluorophosphate, dissolved in D$_2$O (Fig.~\ref{fig:1qexp}(a)), as a two-qubit system, where the $^{19}$F nuclear spin serves as the probe qubit, and the $^{31}$P nuclear spin acts as the system qubit. All experiments were conducted on a $500$ MHz Bruker NMR spectrometer at an ambient temperature of $300$ K. The rotating frame Hamiltonian for the system comprises an internal component and an RF drive, given by $H^\mathrm{FP} = H^\mathrm{FP}_{\mathrm{int}} + H^\mathrm{FP}_{\mathrm{RF}}(t)$. The internal Hamiltonian is expressed as $H^\mathrm{FP}_{\mathrm{int}} = - \omega_\mathrm{F} I_{z}^\mathrm{F} - \omega_\mathrm{P} I_{z}^\mathrm{P} + 2\pi J_\mathrm{FP} I_{z}^\mathrm{F} I_{z}^\mathrm{P},$
while the RF drive Hamiltonian is $H^\mathrm{FP}_{\mathrm{RF}}(t) = \Omega_\mathrm{F}(t) I_{x}^\mathrm{F} + \Omega_\mathrm{P}(t) I_{x}^\mathrm{P}.$ Here, $I_{x,y,z}$ are the spin operators, $(\omega_\mathrm{F}, \omega_\mathrm{P})$ are the adjustable frequency offsets for the $^{19}$F and $^{31}$P spins, and $(\Omega_\mathrm{F}(t), \Omega_\mathrm{P}(t))$ represent the time-dependent RF amplitudes. The scalar coupling constant is $J_\mathrm{FP} = 868.0$ Hz.

\begin{figure}[hbt!]
    \centering
\includegraphics[trim={0cm 0cm 0cm 0cm},clip,width=9cm]{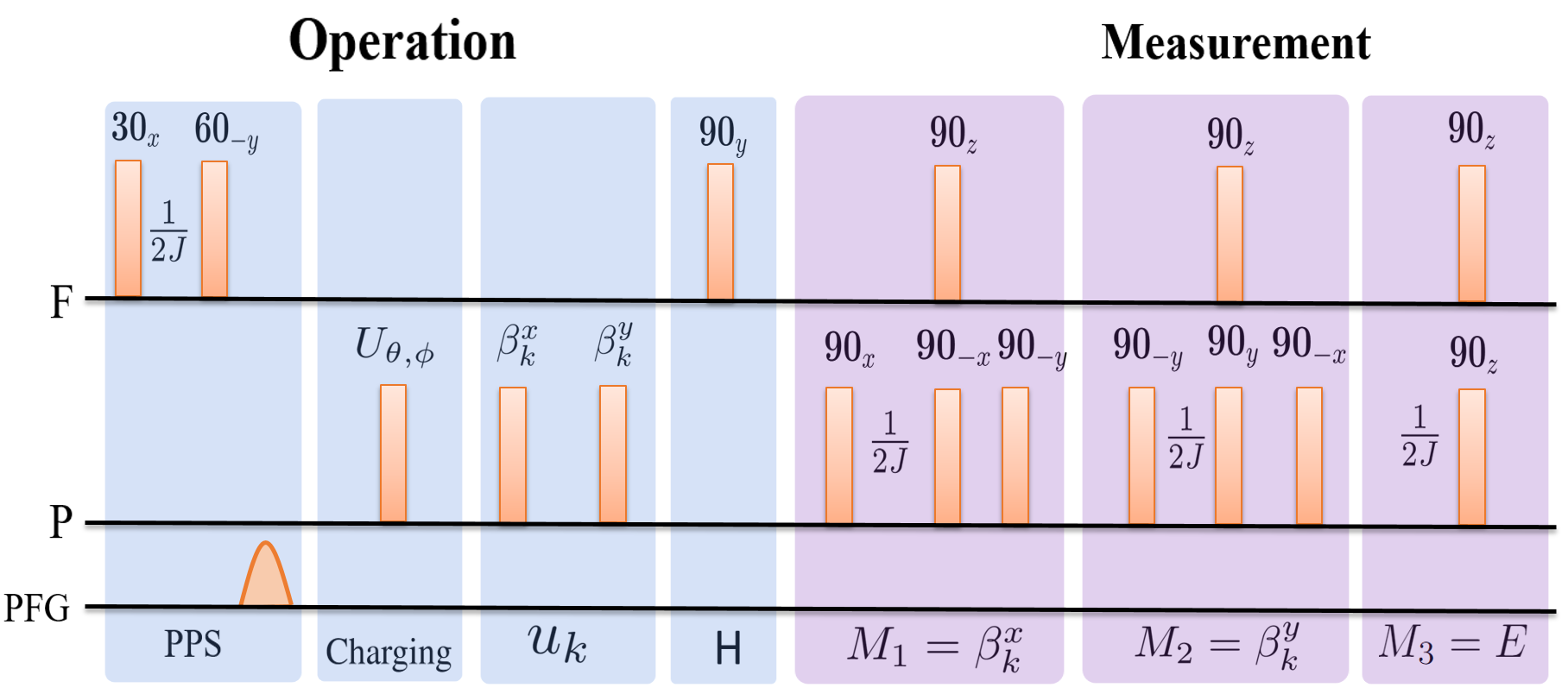}
\caption{NMR pulse sequence for the 1-qubit FQErgo experiment.  From thermal equilibrium state, we first prepare the $\ket{00}$ pseudopure state (PPS) which is rotated by $U_{\theta,\phi} = e^{-i\theta(I_x\cos\phi + I_y\sin\phi)}$ to realize the initial state $\rho_0$.  After FQErgo iterations $u_k = e^{-i\beta^{x}_k I_x\tau}e^{-i\beta^{y}_k I_y\tau}$ on the system qubit $^{31}$P (only one iteration is shown) and Hadamard gate (H) on the probe, we extract (i) one of the drive amplitudes $\beta^{x(y)}$ or (ii) energy $E$. Here the measurement pulse-sequences implement one of  $e^{-i\beta^{x(y)}I_{y(x)}\tau}$ and $e^{-i I_z\tau}$ acting on system $^{31}$P qubit and controlled by $^{19}$F qubit. Finally, $\sigma_{x}$ measurements of $^{19}$F qubit yields drive amplitudes $\beta^{x(y)}$ or  energy $E$.} 
    \label{fig:pulse_1q_2}
\end{figure}

\begin{figure}[hbt!]
\centering
\hspace*{1cm}(a)
\\
\includegraphics[trim={0cm 0cm 0cm 0cm},clip,width=9cm]{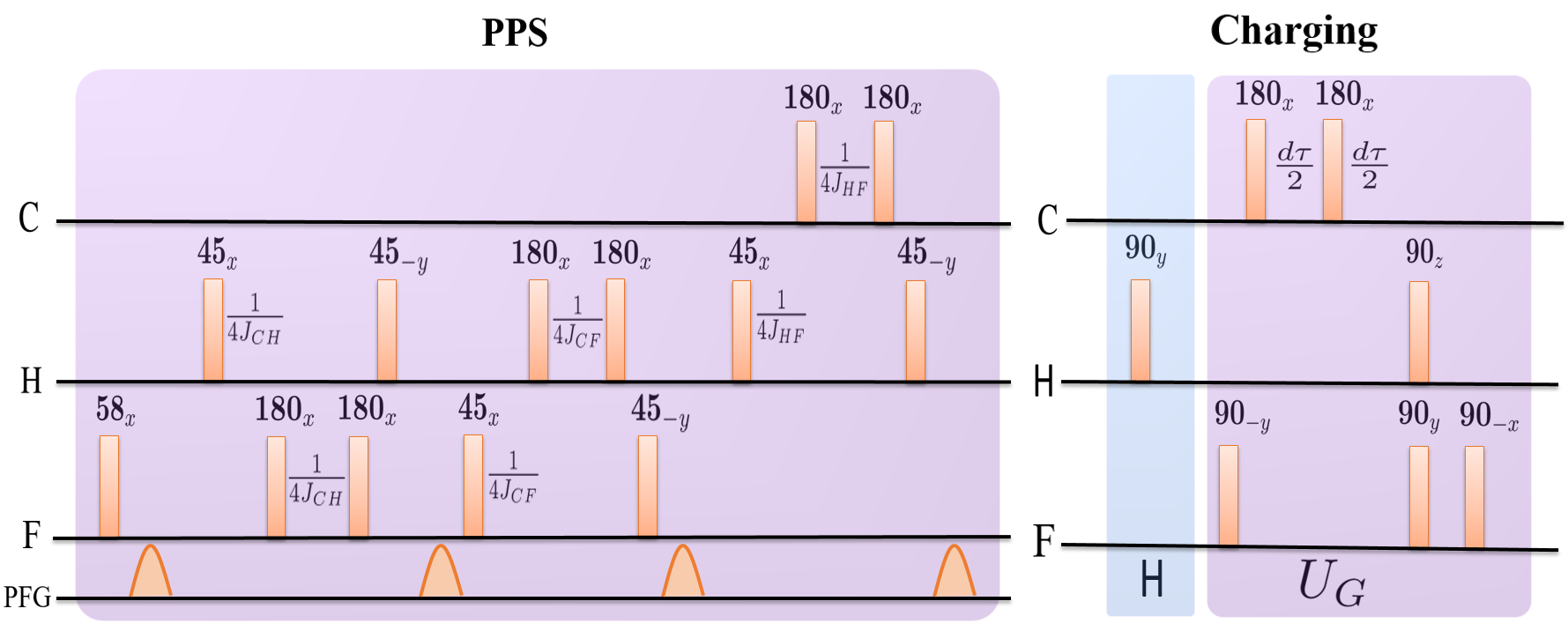}
\\
\hspace*{1cm}(b)\\
\centering
\includegraphics[trim={0cm 0cm 0cm 0cm},clip,width=8cm]{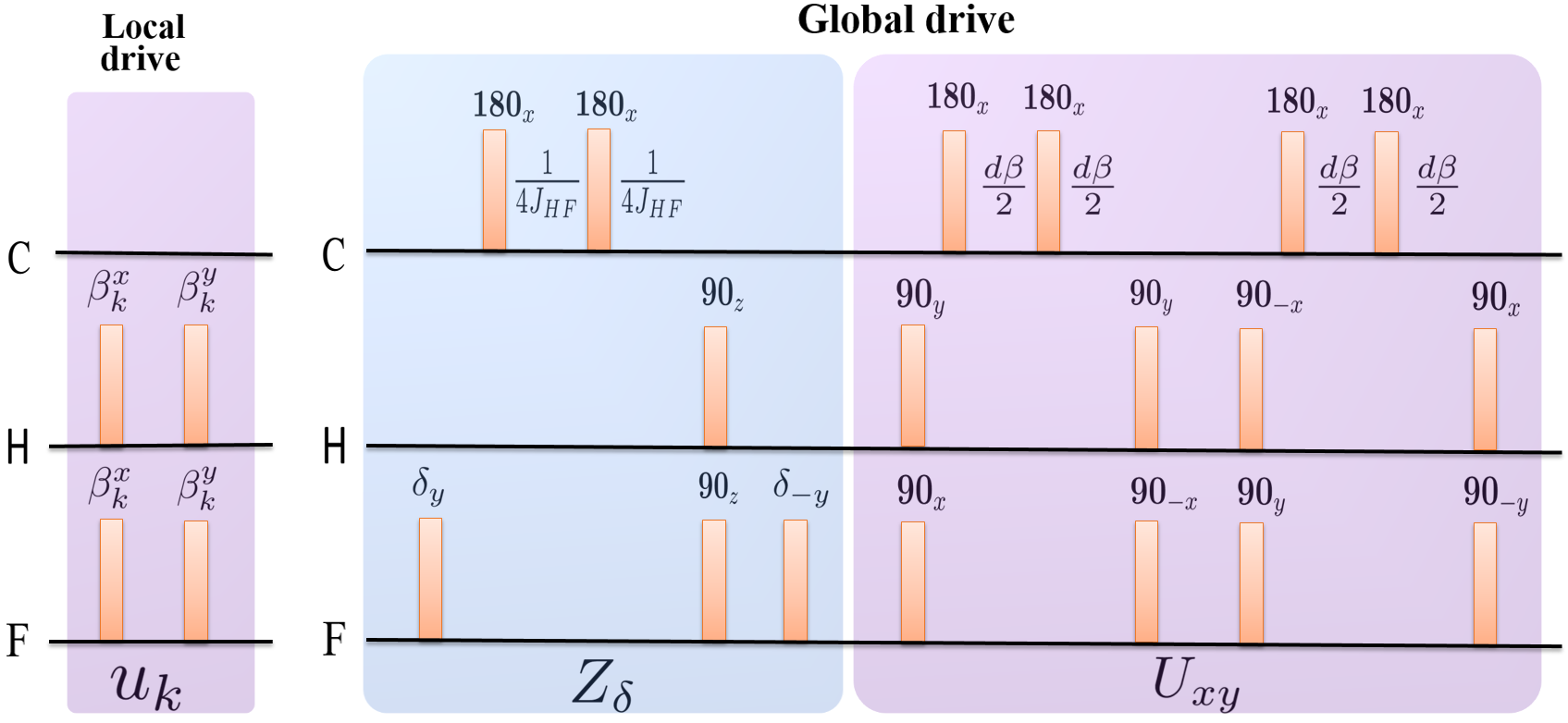} \\
\hspace*{1cm}(c)\\
\includegraphics[trim={0cm 0cm 0cm 0cm},clip,width=9cm]{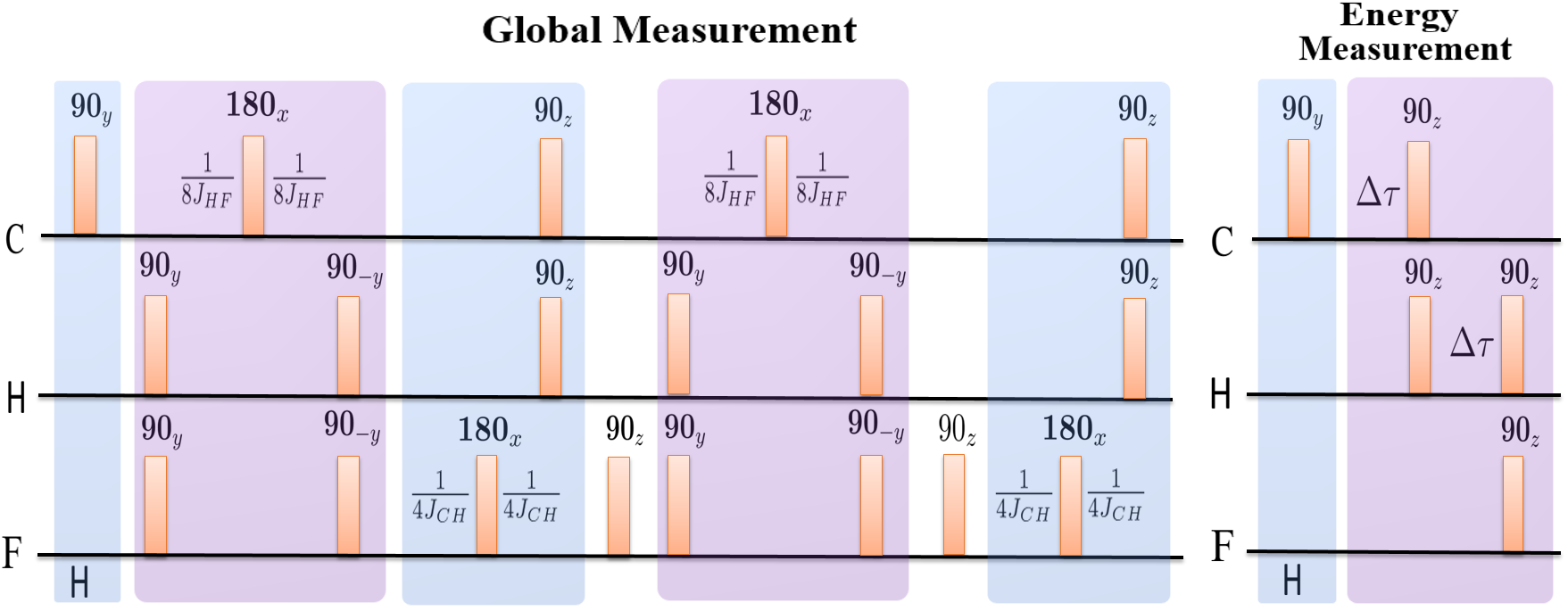}
\caption{
The NMR pulse sequences for 2-qubit ($^1$H,$^{19}$F), FQEergo with one ($^{13}$C) probe qubit: (a) PPS preparation, initializing the system as $\rho^\mathrm{HF}_0$ using entangling unitary $U_G$ after a pseudo-Hadamard on $^1$H. Here delay $d\tau$ is varied to change entanglement entropy. (b) Local and global drives, where $\delta_y = e^{-i\delta I^F_y}$, $U_{xy} = e^{-i\beta^{xy}H_{xy}\tau_{xy}}$ and the delay $d\beta$ controls the strength of the global drive. (c) Extracting global drive strength $\beta^{xy}$ and system energy.}
\label{fig:pulse_2q}
\end{figure}

\begin{figure}
\centering
\includegraphics[trim={0cm 0cm 0.4cm 0cm},clip,width=4cm]{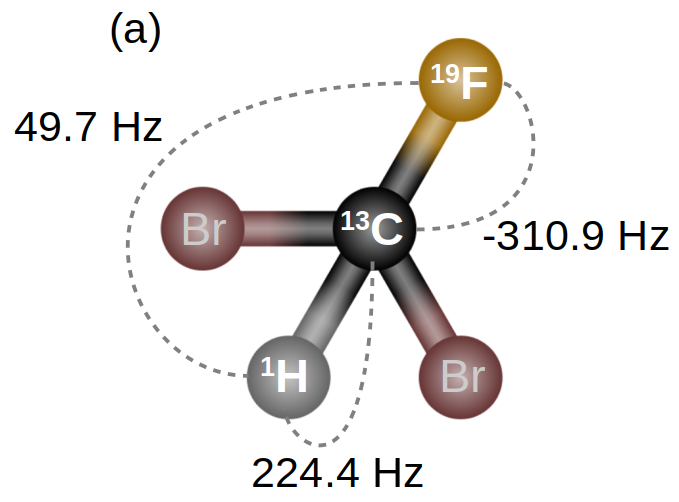}
\includegraphics[trim={1.0cm 0.5cm 0cm 0cm},clip,width=10cm]{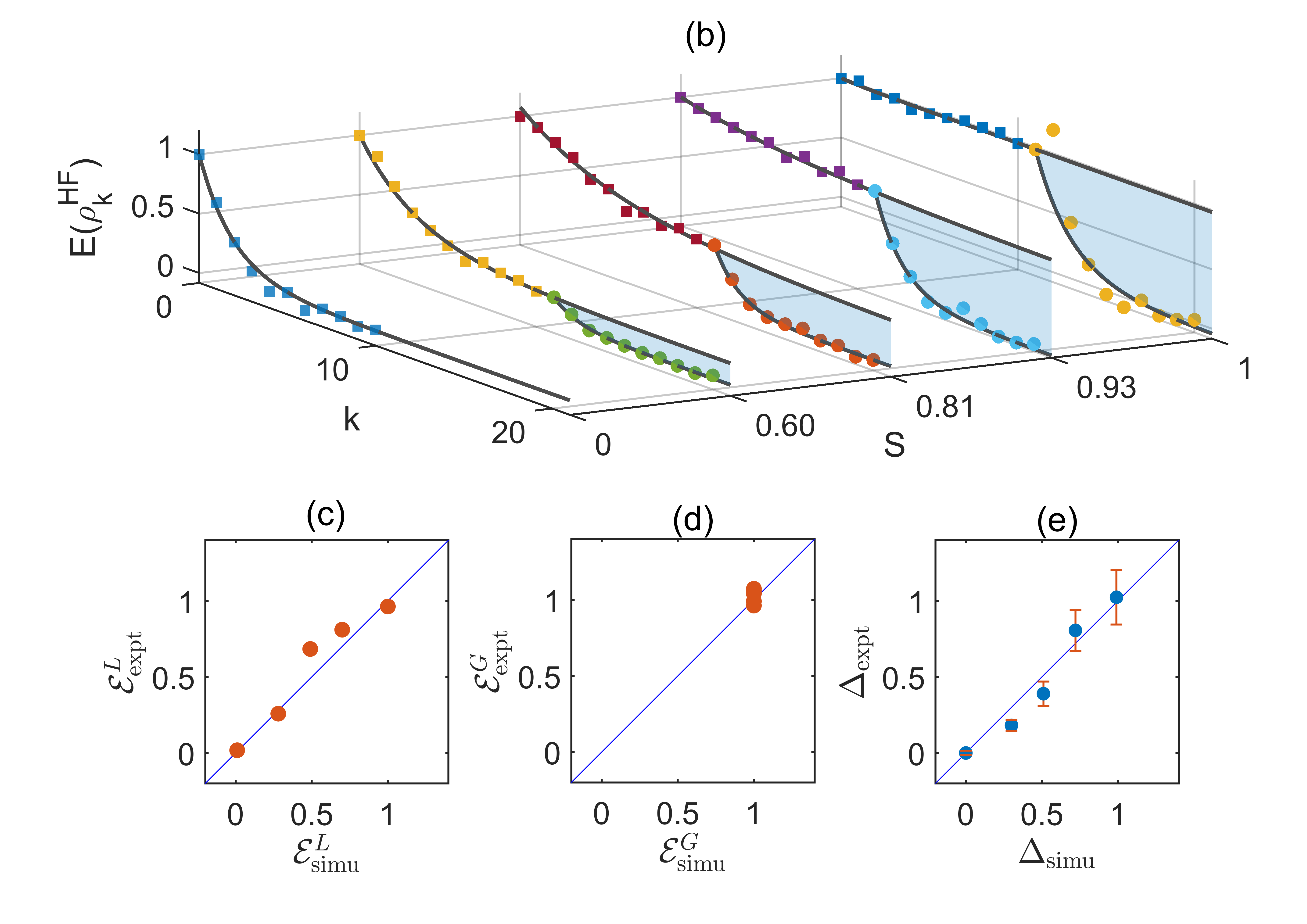}
\caption{\label{fig:2qubitexpt}
(a) Molecular structure of dibromofluoromethane, with various $J$ couplings indicated. 
(b) The experimentally measured two-qubit energy (normalized) vs FQErgo iterations for five initial states with varying entanglement entropy ($S$) as described in the text.  (c,d) Local (c) and global (d) ergotropy experimentally estimated from (b) vs simulated values.
(d) The experimental ergotropy gap obtained from (c,d) vs simulated values.}
\end{figure}

The pulse sequences used for the NMR implementation of FQErgo ass shown in the Fig. \ref{fig:pulse_1q_2}. Starting from the thermal state, we prepare a pseudopure state (PPS) before initializing the system qubit ($^{31}$P) into one of the three initial states $\rho_0$ as depicted in Fig.~\ref{fig:1qexp}. For FQErgo, the initial drives $\beta^x_0 I_x$ and $\beta^y_0 I_y$ are chosen randomly, and their strengths in the $k$th iteration are determined by measuring the ancilla signal $\expv{i[H_0, I_{x(y)}]}{\rho_{k-1}} = \beta^{x(y)}_k$. As shown in Fig.~\ref{fig:1qexp}(b), all three initial states converge to their common passive state within 10 iterations.

The corresponding ergotropy values, shown in Fig.~\ref{fig:1qexp}(c), align closely with numerically simulated results, demonstrating the successful implementation of FQErgo. In the single-qubit case, having measured all three expectation values $\expv{\sigma_{x,y,z}}{\rho_k}$, we reconstruct the density matrix and calculate its state fidelities. The fidelity with the initial state is defined as $F_0(\rho_k) = \tr(\rho_0 \rho_k)$, while the fidelity with the target passive state is $F_p(\rho_k) = \tr(\rho_p \rho_k)$, both evaluated at each iteration. The fidelity profiles, displayed in Fig.~\ref{fig:1qexp}(d), reveal a gradual reduction in the initial state fidelity alongside an increase in the passive state fidelity. In all cases, high-fidelity passive states are achieved.

To estimate two-qubit ergotropy, we utilize a three-spin NMR register consisting of dibromofluoromethane, dissolved in Acetone-D6 (Fig.~\ref{fig:2qubitexpt}(a)). In this setup, $^{13}$C is chosen as the probe qubit, while $^{1}$H and $^{19}$F serve as the system qubits. The rotating frame Hamiltonian includes an internal component and an RF drive, expressed as $H^\mathrm{CHF} = H^\mathrm{CHF}_{\mathrm{int}} + H^\mathrm{CHF}_{\mathrm{RF}}(t)$. The internal Hamiltonian is $H^\mathrm{CHF}_{\mathrm{int}} = - \omega_\mathrm{C} I_{z}^\mathrm{C} - \omega_\mathrm{H} I_{z}^\mathrm{H} - \omega_\mathrm{F} I_{z}^\mathrm{F} 
+ 2\pi J_\mathrm{CH} I_{z}^\mathrm{C} I_{z}^\mathrm{H} 
+ 2\pi J_\mathrm{CF} I_{z}^\mathrm{C} I_{z}^\mathrm{F} 
+ 2\pi J_\mathrm{HF} I_{z}^\mathrm{H} I_{z}^\mathrm{F},$ and the RF drive Hamiltonian is $H^\mathrm{CHF}_{\mathrm{RF}}(t) = \Omega_\mathrm{C}(t) I_{x}^\mathrm{C} + \Omega_\mathrm{H}(t) I_{x}^\mathrm{H} + \Omega_\mathrm{F}(t) I_{x}^\mathrm{F}.$ Here, $(\omega_\mathrm{C}, \omega_\mathrm{H}, \omega_\mathrm{F})$ represent the frequency offsets, $(\Omega_\mathrm{C}(t), \Omega_\mathrm{H}(t), \Omega_\mathrm{F}(t))$ denote the RF amplitudes, and $(J_\mathrm{CH}, J_\mathrm{CF}, J_\mathrm{HF})$ are the scalar coupling constants. The values of these coupling constants are provided in Fig.~\ref{fig:2qubitexpt}(a).

To demonstrate the estimation of local and global ergotropy, we prepare five initial states with varying degrees of entanglement between the system qubits $\mathrm{H}$ and $\mathrm{F}$. The complete NMR pulse sequences are detailed in the Fig. \ref{fig:pulse_2q}. Starting with $\proj{000}$, the state is transformed into $\proj{0} \otimes \rho^\mathrm{HF}_0 = \proj{0} \otimes U_G \proj{00} U_G^\dagger$ using a global unitary $U_G$. This unitary consists of a Hadamard operator on $\mathrm{H}$ and a controlled $e^{-i\nu I_x}$ gate on $\mathrm{F}$, where $\nu$ controls the degree of entanglement. This completes the initialization, after which the work extraction via FQErgo begins.

During $k=10$ iterations, work is extracted using only local drives, $H_d = I_{x(y)}^\mathrm{H}$ and $I_{x(y)}^\mathrm{F}$, similar to Fig.~\ref{fig:ergo_twoqubit}(a). This process achieves the first energy minimization corresponding to the local passive state, as shown in Fig.~\ref{fig:2qubitexpt}(c). From the $k=11$ to the $k=20$ iterations, further energy is extracted using the global drive $H_d = I_{x}^\mathrm{H} I_{y}^\mathrm{F} + I_{y}^\mathrm{H} I_{x}^\mathrm{F}$ in conjunction with the $Z_\delta$ gate and local drives, as depicted in Fig.~\ref{fig:ergo_twoqubit}(b). The inclusion of the global drive enables complete work extraction, ultimately taking the system qubits to the second energy minimization corresponding to the global passive state $\rho^p_\mathrm{HF}$.

Figure~\ref{fig:2qubitexpt}(b) depicts the normalized energies for both local and global work extraction, corresponding to initial states with varying entanglement entropy $S = -\tr[\rho^\mathrm{H}_0 \log \rho^\mathrm{H}_0]$, where $\rho^\mathrm{H}_0 = \tr_\mathrm{F}(\rho_0^\mathrm{HF})$. Figures~\ref{fig:2qubitexpt}(c-d) show the experimentally estimated local and global ergotropy values in comparison with the simulated values, while Fig.~\ref{fig:2qubitexpt}(e) illustrates the experimentally determined ergotropy gaps alongside the simulated values. The strong agreement between experimental and simulated results validates the successful implementation of the FQErgo algorithm.

In summary, we have introduced FQErgo, a feedback-based algorithm designed to prepare passive states and quantify the ergotropy—the maximum unitarily extractable energy—of an unknown quantum state. FQErgo is an iterative and efficient algorithm that utilizes a probe qubit to measure specific expectation values, dynamically adjusting drive strengths to optimize energy extraction. The same probe qubit also helps in real-time monitoring of the system’s energy throughout the process. Through numerical simulations on random states in one- and two-qubit systems, we have demonstrated the robustness of FQErgo in passive state preparation and ergotropy estimation, even in the presence of circuit errors. Experimentally, we implemented FQErgo on multiple initial states of two- and three-qubit NMR registers, successfully preparing their passive states and accurately estimating local and global ergotropies, as well as the ergotropy gap.

FQErgo’s scalability and minimal resource requirements make it suitable for large systems without exponential overhead. Its ability to quantify ergotropy without prior knowledge of the state enables applications such as certifying entanglement in unknown states. Additionally, FQErgo offers a practical tool for energy extraction in quantum technologies like quantum batteries, where states often become unknown during operation. Adaptable across quantum architectures, FQErgo holds promise for advancing quantum thermodynamics and related technologies.

\begin{figure}[hbt!] 
\hspace*{1cm}(a)\\
\centering
\includegraphics[trim={-1cm 0cm 0cm 0cm},clip,width=9cm]{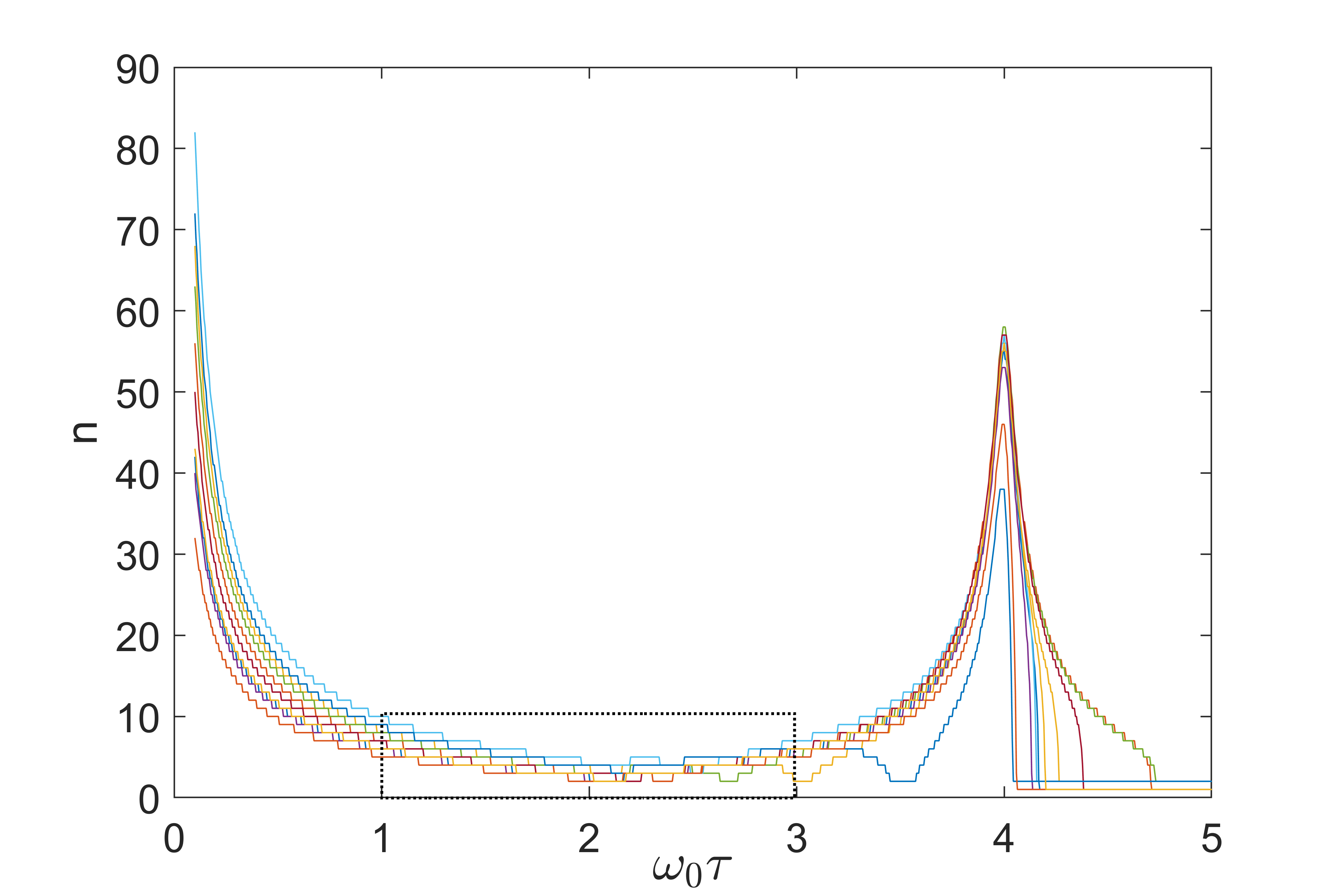}
\includegraphics[trim={0cm -1cm 0cm -1cm},clip,width=9cm]{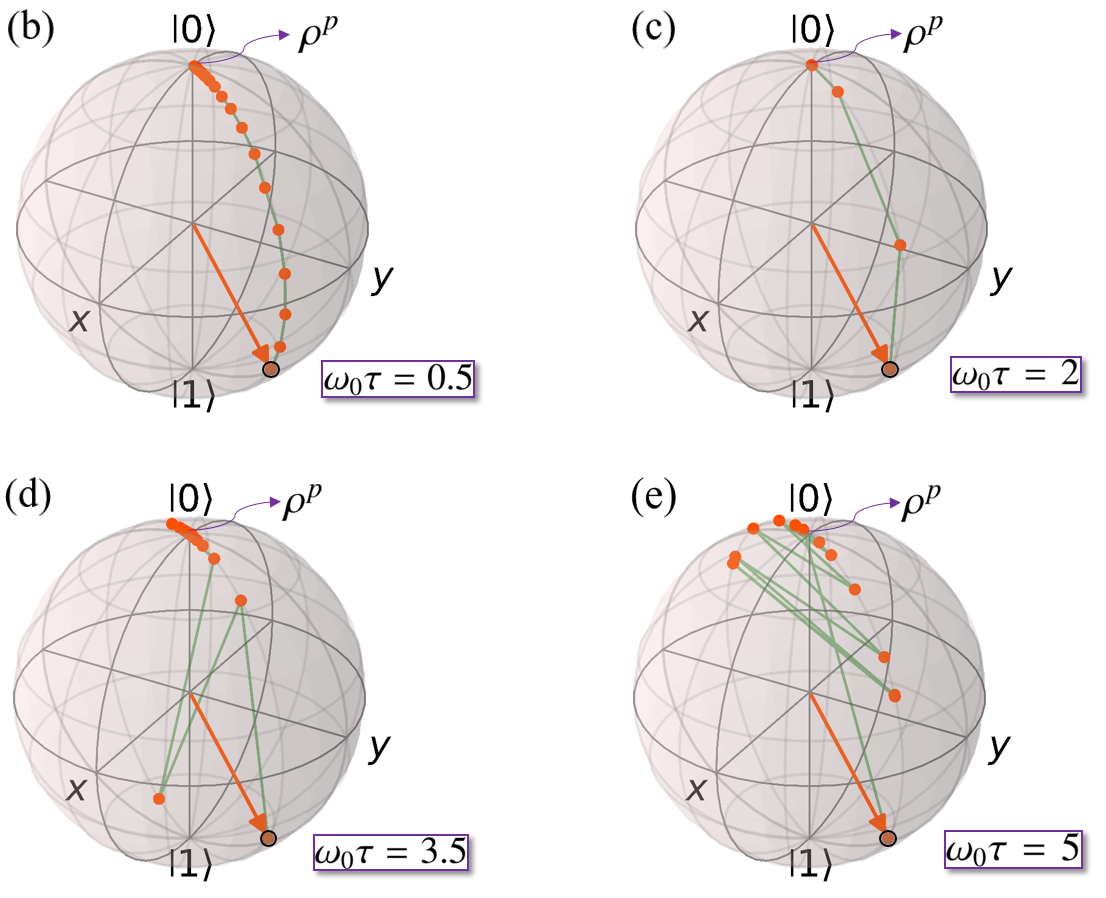}
\hspace*{1cm}(f)\\
\centering
\includegraphics[trim={-1cm 0cm 0cm 0cm},clip,width=9cm]{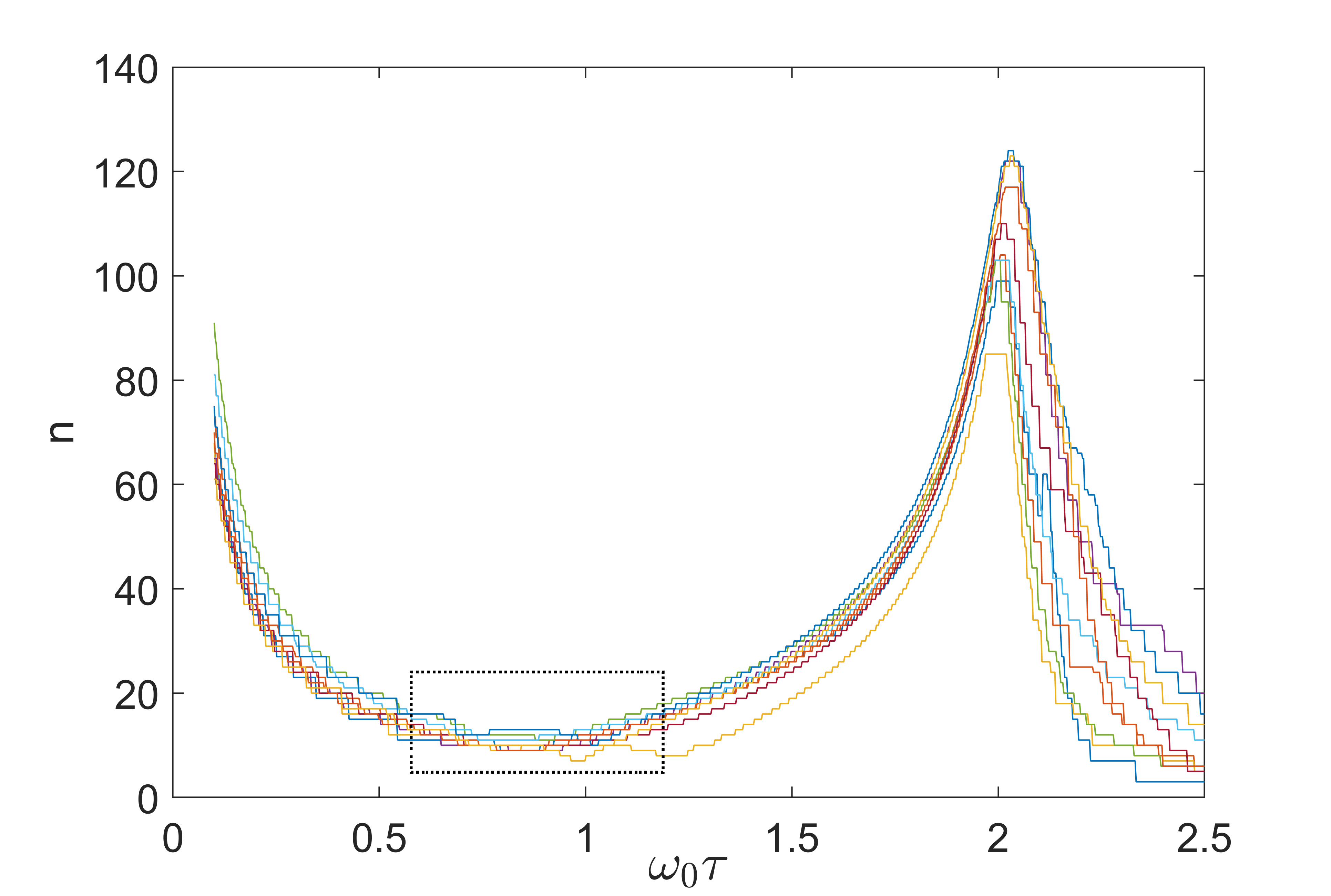}
\caption{\label{fig:speed_1q}
(a) For 10 arbitrary initial states, this plot shows the minimum number of FQErgo iterations $n$ required to reach the passive state in one qubit system v.s. $\omega_0\tau$. (b-e) show Bloch vector evolution from initial state $\ket{\psi} = \cos(\pi/3)\ket{0} + e^{i\pi/3}\sin(\pi/3)\ket{1}$  to passive state $\rho^p$ for different values of $\omega_0\tau$. (f) Same plot as (a) for two qubit system. The area enclosed by dotted rectangles in plots (a) and (f) shows the approximate working regime for $\tau$ values to get the lowest possible FQErgo iteration, i.e., $n$ ranges from $(3,10)$ for one and $(6,20)$ for two qubit systems.}
\end{figure}

{\it Speed analysis:} 
How many feedback-iterations does FQErgo take to reach the passive state from an arbitrary initial state?  
Alternatively, we can ask what should be the optimal step size $\tau$ in $u_k = e^{-i\beta_k \tau H_d }$ that takes the minimum number of iterations $n$ to reach the passive state. 
In the following, we describe numerical 
studies to gain insights into these questions.  
First, let us consider a one qubit system with $\omega_0=1$.
Fig. \ref{fig:speed_1q} (a) shows the number of FQErgo iterations $n$  vs $\omega_0\tau$ for 10 random initial states.  Here we find that the optimal time-step corresponds to $\omega_0\tau$ in the range of 1 to 3.

\begin{acknowledgements}
{\it Acknowledgements:}
Authors acknowledges valuable discussions with Pranav Chandarana, Xi Chen, Mir Alimuddin, Vishal Varma, Arijit Chatterjee, Harikrishnan, and Conan Alexander. J.J. acknowledges support from CSIR
(Council of Scientific and Industrial Research) Project No.
09/936(0259)/2019 EMR - I. We also thank the National Mission on In-
terdisciplinary Cyber-Physical Systems for funding from the DST, Government of India, through the I-HUB Quantum Technology Foundation, IISER-Pune and funding from DST/ICPS/QuST/2019/Q67.   
\end{acknowledgements}

\bibliographystyle{unsrtnat}
\bibliography{ref.bib}

\end{document}